\documentclass[a4paper,fleqn,usenatbib,useAMS]{mnras}

\usepackage{graphicx}	
\usepackage{amsmath}	
\usepackage{amssymb}	
\usepackage{multicol}        
\usepackage{bm}		
\usepackage{pdflscape}	

\newcommand{\kms}{\,km\,s$^{-1}$} 
\newcommand{\calsim}{\textsc{bahamas}}
\newcommand{\planck}{\textit{Planck}}
\newcommand{\wmap}{WMAP9}

\newlength{\colwidth}
\title[Cosmology with velocity dispersions]%
{Cosmology with velocity dispersion counts: an alternative to measuring cluster halo masses}

\author[C.E. Caldwell et al.]{C. E. Caldwell$^1$\thanks{E-mail: C.E.Caldwell@2012.ljmu.ac.uk}, I. G. McCarthy$^1$\thanks{Email: i.g.mccarthy@ljmu.ac.uk}, I. K. Baldry$^1$, C. A. Collins$^1$, J. Schaye$^2$,\newauthor S. Bird$^3$\\
$^1$Astrophysics Research Institute, Liverpool John Moores University, 146 Brownlow Hill, Liverpool, L3 5RF \\
$^{2}$Leiden Observatory, Leiden University, P. O. Box 9513, 2300 RA Leiden, the Netherlands\\
$^{3}$Department of Physics and Astronomy, Johns Hopkins University, 3400 N.~Charles Street, Baltimore, MD 21218, USA}

\date{Version of \today}
\pubyear{2016}

\begin{document}
\label{firstpage}
\pagerange{\pageref{firstpage}--\pageref{lastpage}}
\maketitle

\begin{abstract}
The evolution of galaxy cluster counts is a powerful probe of several fundamental cosmological parameters.  A number of recent studies using this probe have claimed tension with the cosmology preferred by the analysis of the \planck~primary CMB data, in the sense that there are fewer clusters observed than predicted based on the primary CMB cosmology.  One possible resolution to this problem is systematic errors in the absolute halo mass calibration in cluster studies, which is required to convert the standard theoretical prediction (the halo mass function) into counts as a function of the observable (e.g., X-ray luminosity, Sunyaev-Zel'dovich flux, optical richness).  Here we propose an alternative strategy, which is to directly compare predicted and observed cluster counts as a function of the one-dimensional velocity dispersion of the cluster galaxies.  We argue that the velocity dispersion of groups/clusters can be theoretically predicted as robustly as mass but, unlike mass, it can also be directly observed, thus circumventing the main systematic bias in traditional cluster counts studies.  With the aid of the \calsim~suite of cosmological hydrodynamical simulations, we demonstrate the potential of the velocity dispersion counts for discriminating even similar $\Lambda$CDM models.  These predictions can be compared with the results from existing redshift surveys such as the highly-complete Galaxy And Mass Assembly (GAMA) survey, and upcoming wide-field spectroscopic surveys such as the Wide Area Vista Extragalactic Survey (WAVES) and the Dark Energy Survey Instrument (DESI).
\end{abstract}
\begin{keywords}
cosmology: large-scale structure of Universe --- galaxies: clusters --- galaxies: groups --- galaxies: kinematics and dynamics --- neutrinos
\end{keywords}



\section{Introduction}

The abundance of galaxy groups and clusters at a given redshift is directly tied to cosmological parameters that control the growth rate of structure, such as the total matter density ($\Omega_m$), the amplitude of density fluctuations in the early Universe ($\sigma_8$), the spectral index of fluctuations ($n_s$), and the evolution of dark energy (for recent reviews see \citealt{Voit05,Allen11}).  Consequently, measurements of the evolution of the abundance of groups and clusters can be used to constrain the values of these fundamental cosmological parameters.  Recent examples include \citet{Vikhlinin09} and \citet{Bohringer14} using X-ray emission observed with ROSAT, \citet{Benson13} and \citet{Planck13b} using the Sunyaev-Zel'doich (SZ) effect observed with the South Pole Telescope (SPT) and \planck, respectively, and \citet{Rozo10} using the optical maxBCG sample from the Sloan Digital Sky Survey (SDSS).  Upcoming X-ray (eROSITA), SZ (e.g., SPT-3G, ACTpol), and optical (e.g., the Dark Energy Survey, the Large Synoptic Survey Telescope, and Euclid) missions promise to provide even richer datasets that will further enhance this field of study.

In order to compare the observed abundances of groups and clusters with theoretical predictions for a given cosmology, the relation between the observable (e.g., X-ray luminosity, optical richness, weak lensing signal, SZ flux, etc.) and the total mass, including its evolution and scatter, is required to convert the standard theoretical prediction (i.e., the halo mass function) into a prediction for the number counts as a function of the observable.  (A separate important issue is that the predictions normally correspond to the total mass in a dark-matter-only model, but the masses of real groups and clusters can be modified significantly by baryonic physics; e.g., \citealt{Cui14,Velliscig14}.)  One can attempt to determine this observable--mass relation either empirically or by using self-consistent cosmological hydrodynamical simulations.  

However, both methods have their shortcomings.  The empirical route is limited by non-negligible systematic errors in all current methods of total mass estimation (e.g., \citealt{Rozo14}) and can, in any case, generally only be applied to relatively small (generally low-$z$) samples where the data quality is sufficiently high to attempt mass measurement.  The basic problem for the simulation route is that many observable quantities (such as the X-ray luminosity, SZ flux, total stellar mass, etc.) cannot be robustly predicted due to the sensitivity to uncertain `subgrid' physics \citep{LeBrun14}.  

The issue of absolute mass calibration has been brought to the forefront by the \planck~number counts discrepancy \citep{PlanckSZ13}.  Specifically, the best-fit $\Lambda$CDM model based on analyses of the primary CMB data over-predicts the observed number counts by a factor of several (\citealt{Planck13b,Planck15}, see also \citealt{Bohringer14}).  One possible explanation for this discrepancy is the presence of a large `hydrostatic mass bias', such that the adopted X-ray-based masses under-predict the true mass by up to $\sim50\%$ (e.g., \citealt{vonderLinden14}).  Alternatively, there may be remaining relevant systematics in the \planck~CMB data analysis (see, e.g., \citealt{Spergel15,Addison15}), or the discrepancy could be signaling interesting new physics which suppresses the growth of large-scale structure compared to that predicted by a $\Lambda$CDM with parameters fixed (mainly) by the primary CMB at redshift $z\sim1100$, such as free streaming by massive neutrinos (e.g., \citealt{Wyman14,Battye14,Beutler14}).  Clearly, before we can arrive at the conclusion that there is interesting new physics at play, we must rule out the `mass bias' scenario.

One way to independently check the robustness of the claimed discrepancy is to measure the abundance of groups/clusters as a function of some other property that can be theoretically predicted as robustly as mass.  Fortunately, such a variable exists: the velocity dispersion of orbiting satellite galaxies.  The velocity dispersion of the satellites is set by the depth of the potential well and, when in equilibrium, can be expressed via the Jeans equation as:
\begin{equation}
\frac{d[\sigma_{\rm 3D}(r)^2 \rho_{\rm gal}(r)]}{dr} = - \frac{G M_{\rm tot}(r) \rho_{\rm gal}(r)} {r^2},
\end{equation}
\noindent where $\sigma_{\rm 3D}(r)$ is the 3D velocity dispersion profile, $\rho_{\rm gal}(r)$ is the density distribution of the tracer (satellite) population, and $M_{\rm tot}(r)$ is the total mass profile.  Provided the simulations have the correct spatial distribution of tracers (which we discuss further below), they ought to predict the velocity dispersion of satellites as robustly as the mass distribution.  

In practice we do not need to solve the Jeans equations, because the simulations evolve the equations of gravity and hydrodynamics self-consistently, which is necessary given the non-linear complexity of real clusters (e.g., mergers, substructure, asphericity, derivations from equilibrium), and we can directly compare the predicted and observed velocity dispersions.  In particular, in the present study we use the \calsim~suite of simulations, presented in \citet{McCarthy16} (hereafter M16).  These authors calibrated the stellar and AGN feedback models to reproduce the observed local galaxy stellar mass function and the hot gas mass fractions of X-ray groups and clusters.  They then demonstrated that the simulations reproduce a very wide range of other independent observations, including (particularly relevant for the present study) the overall clustering of galaxies (the stellar mass autocorrelation function) and the spatial and kinematic properties of satellites around groups and clusters.

In the present study, we examine the cosmology dependence of the velocity dispersion function and the velocity dispersion counts using \calsim.  We demonstrate that there is a strong dependence, similar to that of the halo mass function, but with the important advantage that the velocity dispersion counts can be directly measured (\textsection\ref{sec:VDF}).  We then propose and verify a simple method for quickly predicting (i.e., without the need to re-run large simulations) the velocity dispersion counts for a given set of cosmological parameters. This method involves convolving the simulated velocity dispersion--halo mass relation (including both intrinsic and statistical scatter and evolution) with the halo mass function predicted for those cosmological parameters (\textsection\ref{sec:model}). We also demonstrate the constraining power of this method for current and future spectroscopic surveys of groups and clusters (\textsection\ref{sec:constraints}). In an upcoming paper, Caldwell et al., in prep, we will apply the theoretical method described in this paper, to the GAMA survey \citep{Driver11, Robotham11} to constrain values of the standard 6-parameter cosmological model.

\section{Simulations}
We use the \calsim~suite of cosmological smoothed particle hydrodynamics (SPH) simulations, which are described in detail in M16.  The \calsim~suite consists of large-volume, $400 \ h^{-1} \ {\rm Mpc}$ on a side, periodic box hydrodynamical simulations.   Updated initial conditions based on the maximum-likelihood cosmological parameters derived from the \wmap~data \citep{Hinshaw13} \{$\Omega_{m}$, $\Omega_{b}$, $\Omega_{\Lambda}$, $\sigma_{8}$, $n_{s}$, $h$\} = \{0.2793, 0.0463, 0.7207, 0.821, 0.972, 0.700\} and the \planck~2013 data \citep{Planck13a} = \{0.3175, 0.0490, 0.6825, 0.834, 0.9624, 0.6711\} are used.  

We also use a massive neutrino extension of \calsim~by M16.  Specifically, McCarthy et al.\ have run massive neutrino versions of the \wmap~and \planck~cosmologies for several different choices of the total summed neutrino mass, $M_\nu$, ranging from the minimum mass implied by neutrino oscillation experiments of $\approx 0.06$ eV \citep{Lesgourgues06} up to $0.48$ eV.  When implementing massive neutrinos, all other cosmological parameters are held fixed apart from the matter density due to cold dark matter, which was decreased slightly to maintain a flat model (i.e., so that $\Omega_{\rm b}+\Omega_{\rm cdm}+\Omega_\nu+\Omega_\Lambda=1$), and $\sigma_8$. The parameter $\sigma_8$ characterises the amplitude of linearized $z=0$ matter density fluctuations on $8 h^{-1}$ Mpc scales.  Instead of holding this number fixed, the amplitude of the density fluctuations at the epoch of recombination (as inferred by \wmap~or \planck~data assuming massless neutrinos) is held fixed, in order to retain agreement with observed CMB angular power spectrum.  Note that other possible strategies for implementing neutrinos are possible (e.g., decreasing $\Omega_\Lambda$ instead of $\Omega_{\rm cdm}$) but McCarthy et al.~have found with small test simulations that the precise choice of what is held fixed (apart from the power spectrum amplitude) does not have a large effect on the local cluster population.  What is most important, is the value of $\Omega_\nu$, which is related to $M_\nu$ via the simple relation $\Omega_\nu = M_\nu / (93.14 \ {\rm eV} \ h^2)$ \citep{Lesgourgues06} and ranges from 0.0013 to 0.0105 for our choices of summed neutrino mass.

The Boltzmann code {\small CAMB}\footnote{http://camb.info/} (\citealt{Lewis00}; April 2014 version) was used to compute the transfer functions and a modified version of V. Springel's software package {\small N-GenIC}\footnote{http://www.mpa-garching.mpg.de/gadget/} to make the initial conditions, at a starting redshift of $z=127$. {\small N-GenIC} has been modified by S.\ Bird to include second-order Lagrangian Perturbation Theory (2LPT) corrections and support for massive neutrinos\footnote{https://github.com/sbird/S-GenIC}.  

The runs used here have $2\times1024^{3}$ particles, yielding dark matter and (initial) baryon particle masses for a \wmap~(\planck~2013) massless neutrino cosmology of $\approx3.85\times10^{9}~h^{-1}~\textrm{M}_{\odot}$ ($\approx4.45\times10^{9}~h^{-1}~\textrm{M}_{\odot}$) and $\approx7.66\times10^{8}~h^{-1}~\textrm{M}_{\odot}$ ($\approx8.12\times10^{8}~h^{-1}~\textrm{M}_{\odot}$), respectively.  (The particle masses differ only slightly from this when massive neutrinos are included.)  

The comoving gravitational softening lengths for the baryon and dark matter particles are set to $1/25$ of the initial mean inter-particle spacing but are limited to a maximum physical scale of $4~h^{-1}$ kpc (Plummer equivalent). The switch from a fixed comoving to a fixed proper softening happens at $z = 2.91$.  $N_{\rm ngb} = 48$ neighbours are used for the SPH interpolation and the minimum SPH smoothing length is limited to $0.01$ times the gravitational softening.  

The simulations were run using a version of the Lagrangian TreePM-SPH code \textsc{gadget3} \citep[last described in][]{Springel05}, which was significantly modified to include new subgrid physics as part of the OverWhelmingly Large Simulations project \citep{Schaye10}.  The simulations include prescriptions for star formation \citep{Schaye08}, metal-dependent radiative cooling \citep{Wiersma09a}, stellar evolution, mass loss, and chemical enrichment \citep{Wiersma09b}, a kinetic supernova feedback prescription \citep{DallaVecchia08}, and a model for black hole mergers and accretion and associated AGN feedback \citep{Booth09}.  For runs with massive neutrinos, the semi-linear algorithm developed by \citet{Bird13}, implemented in \textsc{gadget3}, was used. 

\calsim~is a direct descendant of the OWLS and cosmo-OWLS \citep{LeBrun14,McCarthy14} projects, both of which explored the impact of varying the important parameters of the subgrid models on the stellar and hot gas properties of haloes.  These projects demonstrated that many of the predicted observable properties are highly sensitive to the details of the subgrid modelling, particularly the modelling of feedback processes.  The idea behind \calsim~was therefore to calibrate the supernova and AGN feedback models, using the intuition gained from OWLS and cosmo-OWLS, on some key observables.  M16 elected to calibrate the feedback using the local galaxy stellar mass function and the gas mass fractions of groups and clusters, thereby effectively calibrating on the baryonic content of massive haloes (with $M_{\rm tot} \ga 10^{12} \rm M_\odot$).  

For the purposes of the present study, the accuracy of the calibration is not critically important provided an appropriate selection criteria is imposed on the simulation satellite population; i.e., as long as simulated satellites with {\it total} masses similar to those of the observed satellites are selected (i.e., we want to select the same tracer populations).  In the case of simulations that reproduce the observed galaxy stellar mass function, one can just select simulated galaxies based on their stellar mass (or absolute magnitude).  For simulations that significantly violate the galaxy stellar mass function, and will therefore have an unrealistic mapping between stellar mass and halo mass, one could instead use semi-empirical constraints (e.g., subhalo abundance matching) to re-assign the stellar masses of the simulated galaxies, thereby imposing a realistic mapping between stellar mass and halo mass.  We explicitly demonstrate the lack of sensitivity of the velocity dispersions to the details of the subgrid modelling in Section \ref{sec:subgrid}.

\section{Cosmology dependence of velocity dispersion counts}

In this section we compute the one-dimensional velocity dispersion counts from the simulations, demonstrating that they exhibit a strong cosmology dependence, similar to that of the cluster mass counts.  We first specify how we estimate the velocity dispersion of simulated groups and clusters.

\subsection{Galaxy and group selection criteria}
\label{sec:selection}

Before we can calculate velocity dispersions for the simulated groups and clusters, an appropriate tracer population must be selected.  Previous studies (usually based on N-body simulations) often selected bound dark matter particles (e.g., \citealt{Evrard08}).  However, the satellite galaxy population could in principle have a different spatial/kinematic distribution compared to the underlying smooth dark matter distribution, e.g., through the effects of dynamical friction, or just simply differences in the time of accretion of satellites compared to that of the (smooth) dark matter component.  Indeed, many previous studies have found that the satellites are more spatially-extended (i.e., have a lower concentration) than what is measured for the total mass distribution (e.g., \citealt{Carlberg97,Lin04,Budzynski12,vanderBurg15}). M16 have shown that in the case of \calsim, the satellites have a negative velocity bias (i.e., a lower velocity dispersion) with respect to the underlying dark matter particles.

\begin{figure*}
\includegraphics[width=0.995\columnwidth]{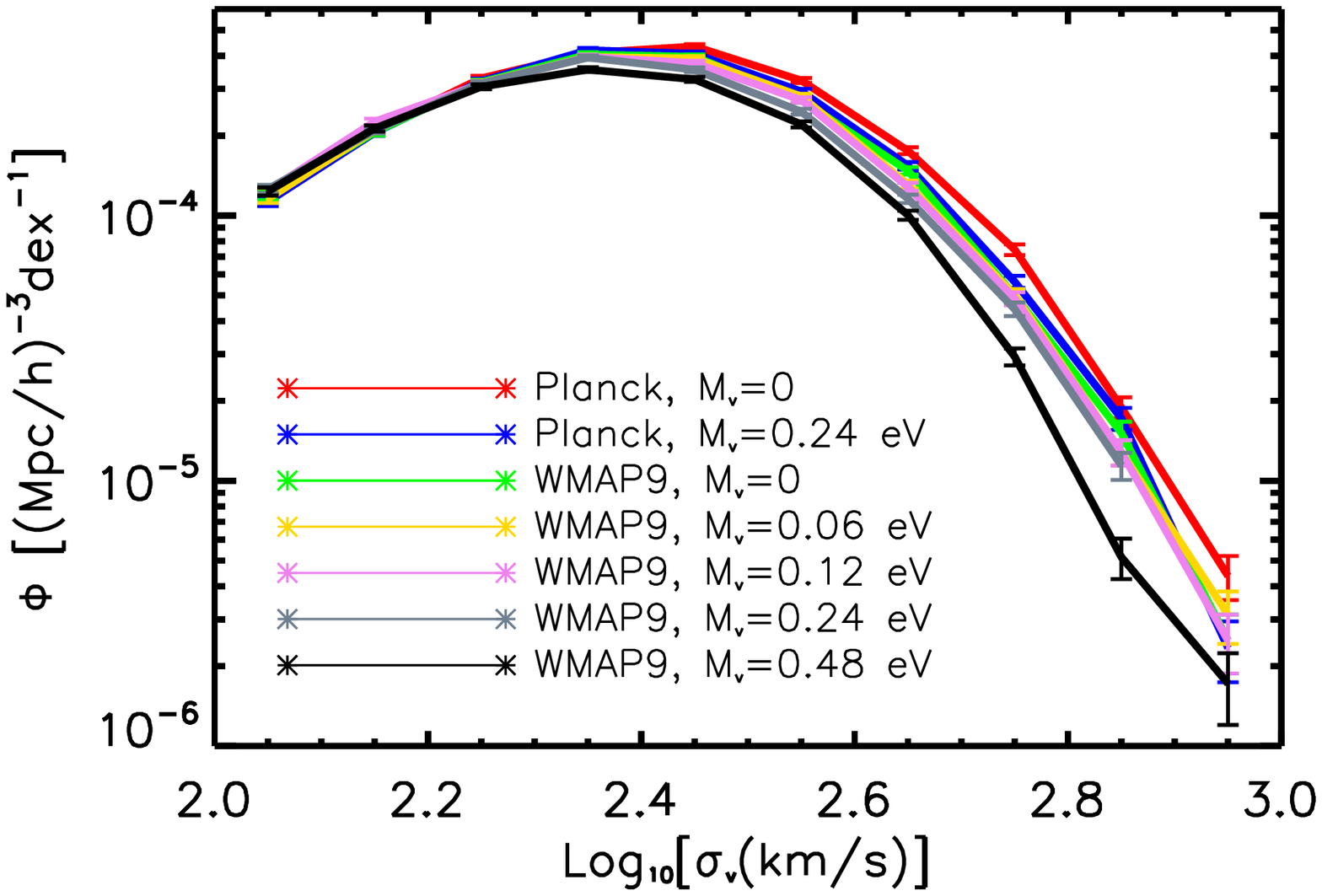}
\includegraphics[width=0.995\columnwidth]{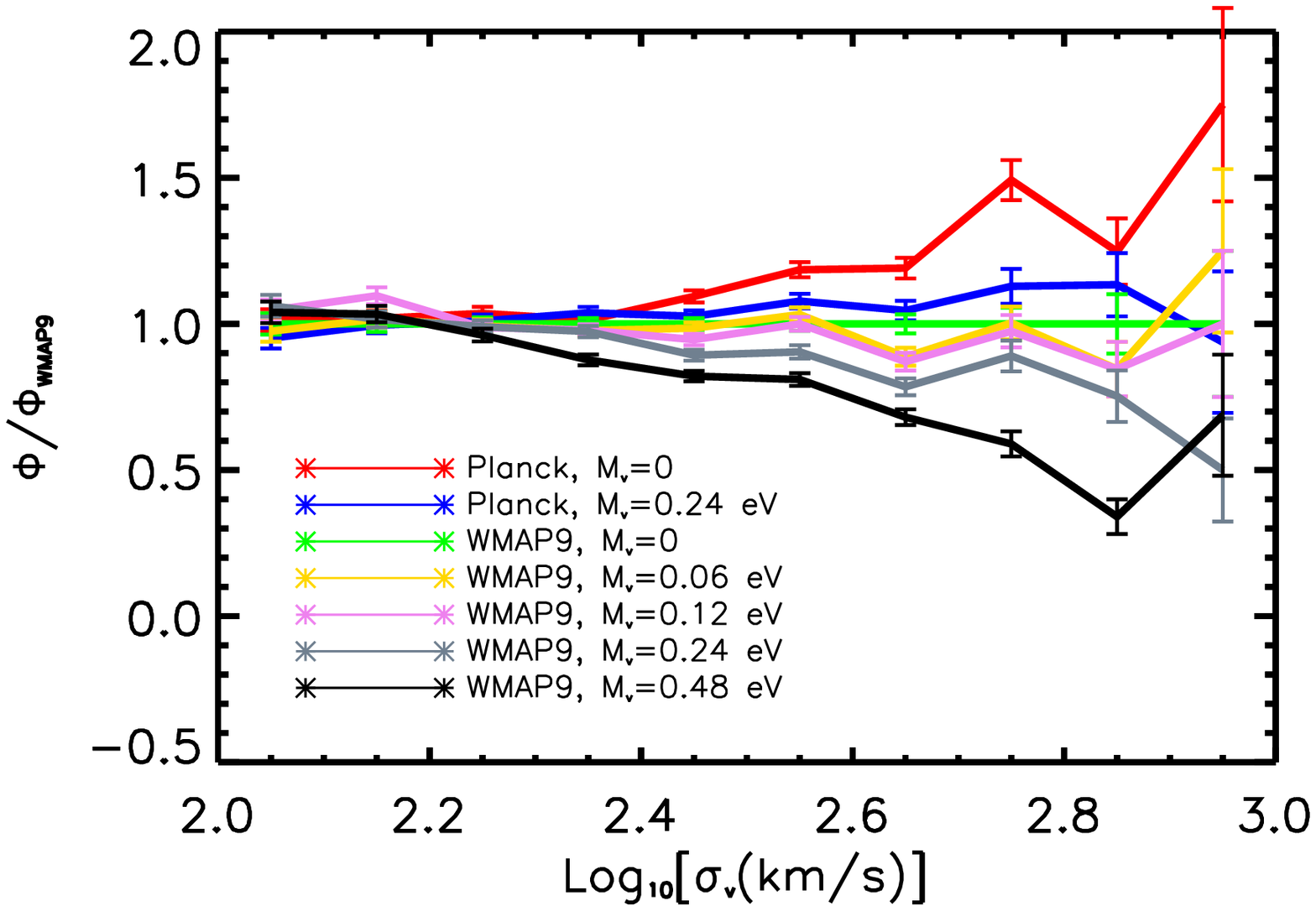}
\caption[]{The predicted one-dimensional velocity dispersion function $\Phi\equiv  \mathrm{d}n/ \mathrm{d}\log_{10}\sigma_v$, or VDF, for the \wmap~and \planck~2013 cosmologies for various choices of neutrino mass (including massless) at $z=0$.  The error bars represent Poisson sampling errors and are estimated as the square root of the number of systems in a given velocity dispersion bin divided by the simulation volume.  The right panel shows the ratio of the predicted VDFs with respect the \wmap~case with massless neutrinos.  Velocity dispersions are calculated using member galaxies within a 3D radius $r_{200m}$  that have stellar masses $M_* \ge 10^{10} \rm M_\odot$.  Only groups/clusters having at least 5 member galaxies are included, which is why the VDFs turns over at $\log_{10}\sigma_v \mathrm{km/s} \la2.4$.  The predicted VDFs are a strong function of cosmology, like the halo mass function, but offer the advantage that velocity dispersions are directly measurable.}
\label{fig:vdf_data}
\end{figure*} 

With cosmological hydrodynamical simulations we can move beyond selecting dark matter particles and identify satellite galaxies.  We define galaxies in the simulations as self-gravitating substructures (identified with SUBFIND algorithm, \citealt{Springel01,Dolag09}) with non-zero stellar mass.  For the analysis below, we present results based on selecting groups of 5 or more satellites with stellar masses exceeding $10^{10} \rm M_\odot$ (i.e., that are `resolved' in the simulations) and that are within a 3D radius $r_{200m}$, which is the radius that encloses a mean density that is 200 times the mean universal density at that redshift [i.e., $200 \Omega_m(z) \rho_{\rm crit}(z)$].  Note that the derived velocity dispersions are not strongly sensitive to these choices, however, owing to the fact that the total mass distribution is fairly close to isothermal and that the radial distribution of satellites is not a strong function of stellar mass (M16).  For completeness, in Appendix A we provide fits to the velocity dispersion--halo mass relation for various choices of mass definition and aperture (including both spherical and cylindrical radii) for selecting satellites.


\subsection{Velocity dispersion calculation}
\label{sec:gapper}
With a tracer population in hand, we proceed to calculate the velocity dispersions of the simulated groups and clusters.  There are several possible methods for calculating the velocity dispersion of a system (simulated or real), including calculating a simple root-mean-square (RMS) or fitting a normal distribution to the galaxy redshifts.  We have decided to use the so-called `gapper' algorithm \citep{WainerThissen76}, due to its practical application to observations (e.g., \citealt{Eke04, Robotham11, Ruel14, Proctor15}) and robustness at low richness \citep{Beers90}.  With the gapper method, the velocities are sorted from least to greatest and the velocity dispersion is then estimated as:

 \begin{equation}
 \label{eq:gapper}
\sigma_{\rm gap}=\frac{\sqrt\pi}{N(N-1)} \sum^{N-1}_{i=1} w_i g_i ,
\end{equation}
\noindent with $w_i = i(N-i)$ and $ g_i=v_{i+1} - v_i $,
where N is the number of galaxies in the group or cluster, and v$_i$ is the \textit{i}th velocity from a list of the group's galaxies' velocities, that has been sorted in ascending order.

Although, statistically, the gapper method does not require the central object to be removed before calculation of the velocity dispersion, we have found that the mean gapper velocity dispersions of galaxy groups are lower than then mean RMS velocity dispersion with the central removed. This is likely due to the central galaxy moving at a velocity that is not typical of the satellite population.  Therefore we follow \citet{Eke04} and scale $\sigma_{\rm gap}$ up by $[N/(N-1)]^{1/2}$ to account for these effects.  Clearly, this correction is only relevant for low-mass groups with richnesses approaching unity, for which we have found that including this correction results in velocity dispersion estimates that are more stable to changes in the  stellar mass cut used to select satellites.  We use the symbol $\sigma_v$ to denote the gapper velocity dispersion after it has been multiplied by the Eke et al.~correction. 

Although the simulation provides velocities in three dimensions, we limit our analysis to using only one dimension (we do not average the three one-dimensional velocity components) to replicate the information available in real observations.  Therefore, $\sigma_v$ is a 1-dimensional velocity dispersion.

\subsection{Velocity dispersion function and number counts}
\label{sec:VDF}
\subsubsection{Velocity Dispersion Function}

We define the velocity dispersion function ($\Phi$), or VDF, as the number of systems per unit comoving volume per decade in velocity dispersion; i.e., $\Phi\equiv  \mathrm{d}n/ \mathrm{d}\log_{10}\sigma_v$.  In Figure \ref{fig:vdf_data} (left panel) we show the $z=0$ VDFs for various cosmologies. The errors on the VDF are the number of groups in a velocity dispersion bin, divided by the volume of the simulation. The VDF clearly depends on cosmology, as expected.  Note that the turnover in the VDF at low $\sigma_v$ is due to the fact that we impose a richness cut of $N \ge 5$ on our simulated groups (i.e., each system must have at least 5 galaxies meeting the selection criteria noted in Section \ref{sec:selection}).  This is inconsequential for our purposes, since we are primarily interested in the {\it relative} differences between different cosmologies at the moment.

The right panel of Figure \ref{fig:vdf_data} shows the ratios of the predicted VDFs with respect to that of the \wmap~case with massless neutrinos.  It more clearly demonstrates the strong cosmology dependence of the VDF.  For example, at a velocity dispersion $\sigma_v\sim1000$ km/s, adopting a \planck~2013 cosmology results in $\approx50\%$ more systems compared to adopting a \wmap~cosmology (both assuming massless neutrinos).  Even at a relatively modest velocity dispersion of $\sim300$ km/s (corresponding roughly to haloes with masses $\sim10^{14} \rm M_\odot$) the difference is still significant ($\approx20\%$).  The introduction of massive neutrinos suppresses the number of high-velocity dispersion systems, as expected.

\subsubsection{Number counts}

\begin{figure}
\includegraphics[width=0.995\columnwidth]{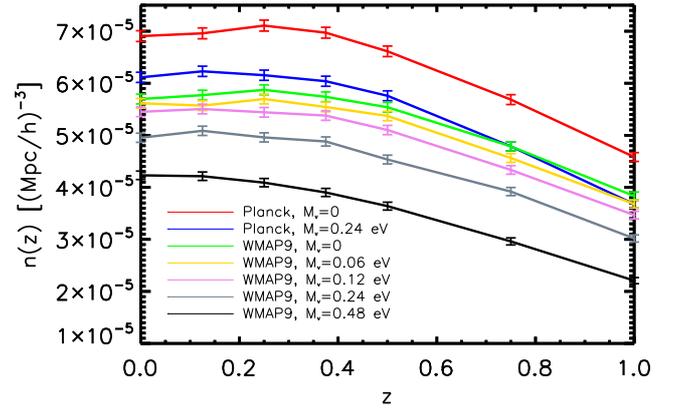}
\caption[]{The number density of systems with $\sigma_v\ge300$ km/s as a function of redshift for the various cosmologies we consider in Figure 1. The error bars are the square root of the number of objects in a redshift bin, divided by the volume of the simulation. }
\label{fig:nz_data}
\end{figure}

Because the systems of interest have space densities of only $< 10^{-4}$ Mpc$^{-3}$, observational surveys covering a large fraction of the sky are required to detect massive systems in appreciable numbers.  Given the limited statistics, splitting the sample into bins to measure a differential function, like the VDF, may not always be possible, particularly as one moves to higher redshifts.  An alternative, therefore, is to measure the cumulative number counts above some threshold value in the observable.  With this in mind, we show in Figure \ref{fig:nz_data} the number density of systems with $\sigma_v\ge300$ km/s as a function of redshift for the various cosmologies we consider.  This plot is analogous to the SZ number counts in \citet{PlanckSZ13} (see their Fig.~7).  There is a clear stratification between the different cosmologies presented in this plot.  

It is interesting to note that the velocity dispersion number counts do not drop off very steeply with redshift, in contrast to the halo mass counts.  This is due to the fact that the radius enclosing a spherical overdensity mass (e.g., $r_{200m}$) decreases with increasing redshift (because the background density increases with increasing redshift), and hence the typical orbital velocity, which scales as $(G M/r)^{1/2}$, will increase for a halo of fixed mass with increasing redshift.  The net result of this is that the number of systems above a given threshold value in velocity dispersion will not drop off as quickly as the number of haloes above a given halo mass threshold.


\section{Predicting the velocity dispersion counts for different cosmologies}
\label{sec:model}

In the previous section, we calculated, directly from the simulation, cluster number counts as a function of velocity dispersion and redshift for seven different combinations of cosmology and neutrino masses. The computational expense of running large simulations like \calsim~prohibits us from running a dense grid of cosmologies for comparison with observations, which is ultimately necessary to determine not only the best-fit cosmology, but also the uncertainties in the best-fit cosmological parameters.  We therefore require a means to rapidly compute the predicted velocity dispersion counts for many different cosmologies.

Here we propose a method to combine the results of the simulations with the halo model formalism to predict the velocity dispersion counts.  Specifically, below we characterise the velocity dispersion--halo mass relation in the simulations (mean relation, scatter, and evolution) with simple functions and show that when convolved with the distribution of halo masses in the simulations, it closely predicts the velocity dispersion counts.  One can therefore take advantage of popular theoretical models for the halo mass function (e.g., \citealt{Sheth01,Tinker08}), provided they are appropriately modified for the effects of baryon physics (e.g., \citealt{Cui14,Velliscig14}), and our velocity dispersion--halo mass relation to quickly and accurately predict the velocity dispersion counts as a function of cosmological parameters.


\subsection{Velocity dispersion--halo mass relation}
\label{sec:powerlaw}

\begin{figure}
\includegraphics[width=0.995\columnwidth]{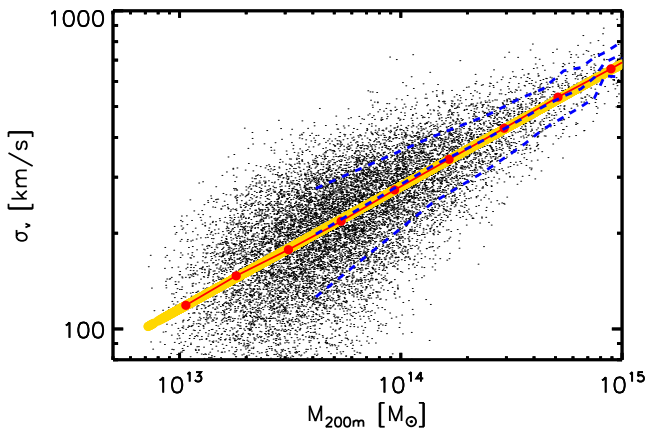}
\caption[]{The velocity dispersion--halo mass relation for the \planck~2013 cosmology with massless neutrinos.  Velocity dispersions are calculated using member galaxies within a 3D radius $r_{200m}$ and that have stellar masses $M_* \ge 10^{10} \rm M_\odot$.   The small black dots show the individual groups and clusters, the red circles connected by a solid red curve show the mean velocity dispersions in halo mass bins, and the gold line represents the best-fit power-law to the mean relation (i.e., to the red circles).  The upper and lower dashed blue curves enclose 68\% of the population.  The mean relation and scatter are well-represented by a simple power-law relation with lognormal scatter. }
\label{fig:msmatch}
\end{figure}

\subsubsection{Present-day relation}

We model the mean relation between velocity dispersion and halo mass at a given redshift using a simple power-law of the form:
\begin{equation}
\label{eq:ms}
\langle \sigma_v | M_\Delta \rangle = a \Big(\frac{M_\Delta}{10^{14} \rm M_\odot}\Big)^{b} .
\end{equation}
\noindent To derive the mean relation, we first compute the mean velocity dispersions in mass bins of width 0.25 dex.  A power-law is then fit to these mean velocity dispersions.  Note that by deriving the mean velocity dispersion in bins of halo mass before fitting the power-law, we are giving equal weight to each of the mass bins.  If instead one were to fit a power-law to all systems, groups would clearly dominate the fit due to their much higher abundance compared to clusters.  However, we want to accurately characterise the relation over as wide a range of halo masses as possible, motivating us to bin the data in terms of mass first.

In Figure \ref{fig:msmatch} we show the velocity dispersion--halo mass relation for the \planck~2013 cosmology (with massless neutrinos).  The small black dots show the individual groups and clusters, the red circles connected by a solid red curve show the mean velocity dispersions in halo mass bins, and the gold line represents the best-fit power-law to the mean relation (i.e., to the red circles).

The mean $z=0$ $\sigma_v$--halo mass relation for this particular \planck~2013 cosmology simulation, adopting a group mass defined as $M_{200m}$, and selecting satellites within $r_{200m}$ with a minimum stellar mass of $10^{10}$ M$_\odot$, is:
\begin{equation}
\langle \sigma_v | M_{200m} \rangle_{z=0} = 280.5 \pm{1.0}\ {\rm km/s} \ \Big(\frac{M_{200m}}{10^{14} \rm M_\odot}\Big)^{0.385 \pm{0.003}} .
\end{equation}

Note that although this relation was derived from simulations run in a \planck~2013 cosmology, the best-fit relations for other cosmologies we have examined are virtually identical.  This likely just reflects the fact that once systems are virialized, the orbital motions of satellites are mainly sensitive to the {\it present} potential well depth and not to how that potential well was assembled (which will change with the  cosmology).  The lack of a cosmological dependence of the velocity dispersion--halo mass relation, at redshifts less than one, considerably simplifies matters, as it means one does not need to re-fit the relation for every cosmology and can just convolve this `universal' relation with the halo mass function (which does depend strongly on cosmology, but for which there are many models in the literature for quickly calculating the HMF for a particular choice of cosmological parameters).

It is interesting to note that the best-fit relation has a slope of $b=0.385$, which is comparable to the self-similar prediction of $1/3$.  A similar finding has been reported recently by \citet{Munari13}, who also used cosmological hydro simulations to examine the velocity dispersion--halo mass relation (although they did not address the issue of velocity dispersion counts).  

Furthermore, the best-fit {\it amplitude} differs significantly from that found previously by \citet{Evrard08} for dark matter particles in pure N-body cosmological simulations:
\begin{equation}
\langle \sigma_v | M_{200m} \rangle_{{\rm Evrard+08}, z=0} = 342 \pm{1} \ {\rm km/s} \ \Big(\frac{M_{200m}}{10^{14} \rm M_\odot}\Big)^{0.355 \pm{0.002}} ,
\end{equation}

\noindent suggesting that the satellite galaxies have a $\approx -20\%$ velocity bias with respect to the velocity dispersion of the dark matter.  M16 have confirmed this to be the case for \calsim~by comparing the satellite velocity dispersions to the dark matter particles in the same simulation.  

Is the mass--velocity disperion relation derived from \calsim~realistic?  As we have already argued, self-consistent simulations ought to be able to predict velocity dispersions as reliably as they can halo masses, so long as an appropriate selection is applied.  However, one can also attempt to check the realism of the relation by comparing to observational constraints, noting the important caveat that observational halo mass estimates could have relevant systematic biases (which is what motivated our proposed use of velocity dispersion counts in the first place).  Of the methods currently in use to estimate halo masses, weak lensing mass reconstructions are expected to have the smallest bias (of only a few percent) when averaged over a large number of systems (e.g., \citealt{Becker11,Bahe12}).  M16 have compared the mean halo mass--velocity dispersion relation from \calsim\ (using the same galaxy stellar mass selection as our fiducial selection employed here) to that derived from the maxBCG cluster sample \citep{Koester07}, derived by combining the stacked velocity dispersion--richness relation of \citet{Becker07} with the stacked weak lensing mass--richness relation of \citet{Rozo09}.  Fig.\ 10 of M16 demonstrates the excellent agreement between the simulations and the observational constraints.

For completeness, in Table \ref{table:coeffs} of Appendix A we provide the best-fit power-law coefficients for the mean velocity dispersion--halo mass relation for different combinations of mass definition and aperture.

\begin{figure}
\includegraphics[width=0.995\columnwidth]{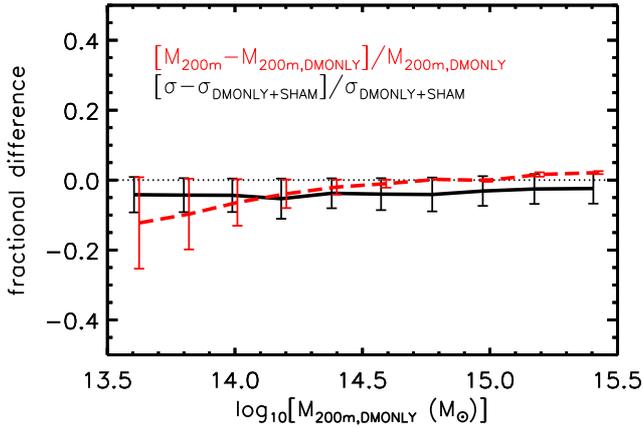}\caption[]{Mean fractional differences in the velocity dispersion and halo mass of matched haloes between \calsim~and a corresponding dark matter-only simulation (WMAP9 cosmology).  The error bars represent the standard error on the mean.  Note that we use subhalo abundance matching (SHAM) to assign stellar masses to subhaloes in the dark matter only simulations (see text), in order to apply the same selection criteria as imposed on the hydro simulations.  Baryon physics (AGN feedback, in particular) lowers the halo masses of galaxy groups by $\sim10\%$ (consistent with \citealt{Velliscig14}) and also reduces the velocity dispersions by $\approx5\%$.}
\label{fig:simcomp}
\end{figure}

\subsubsection{Sensitivity to baryon physics}
\label{sec:subgrid}
As discussed in Section 1, predictions for the internal properties of groups and clusters (particularly of the gaseous and stellar components) are often sensitive to the details of the subgrid modelling of important feedback processes.  One can attempt to mitigate this sensitivity by calibrating the feedback model against particular observables, as done in \calsim.  We anticipate that the velocity dispersions of satellites will be less sensitive to the effects of feedback than, for example, the gas-phase properties or the integrated stellar mass, since the dynamics of the satellite system is driven by the depth of the potential well which is dominated by dark matter.  However, the total mass (dark matter included) of groups and clusters can also be affected at up to the 20\% level with respect to a dark matter only simulation, if the feedback is sufficiently energetic (e.g., \citealt{Velliscig14}).  The feedback will also reduce the masses of the satellites prior to accretion.  The reduction of the satellite and host masses could in turn also affect the resulting spatial distribution of the satellites somewhat, and hence the velocity dispersion.  Given these potential effects, it is therefore worth explicitly testing the sensitivity of the velocity dispersions to baryon physics.

\begin{figure*}
\includegraphics[width=0.995\columnwidth]{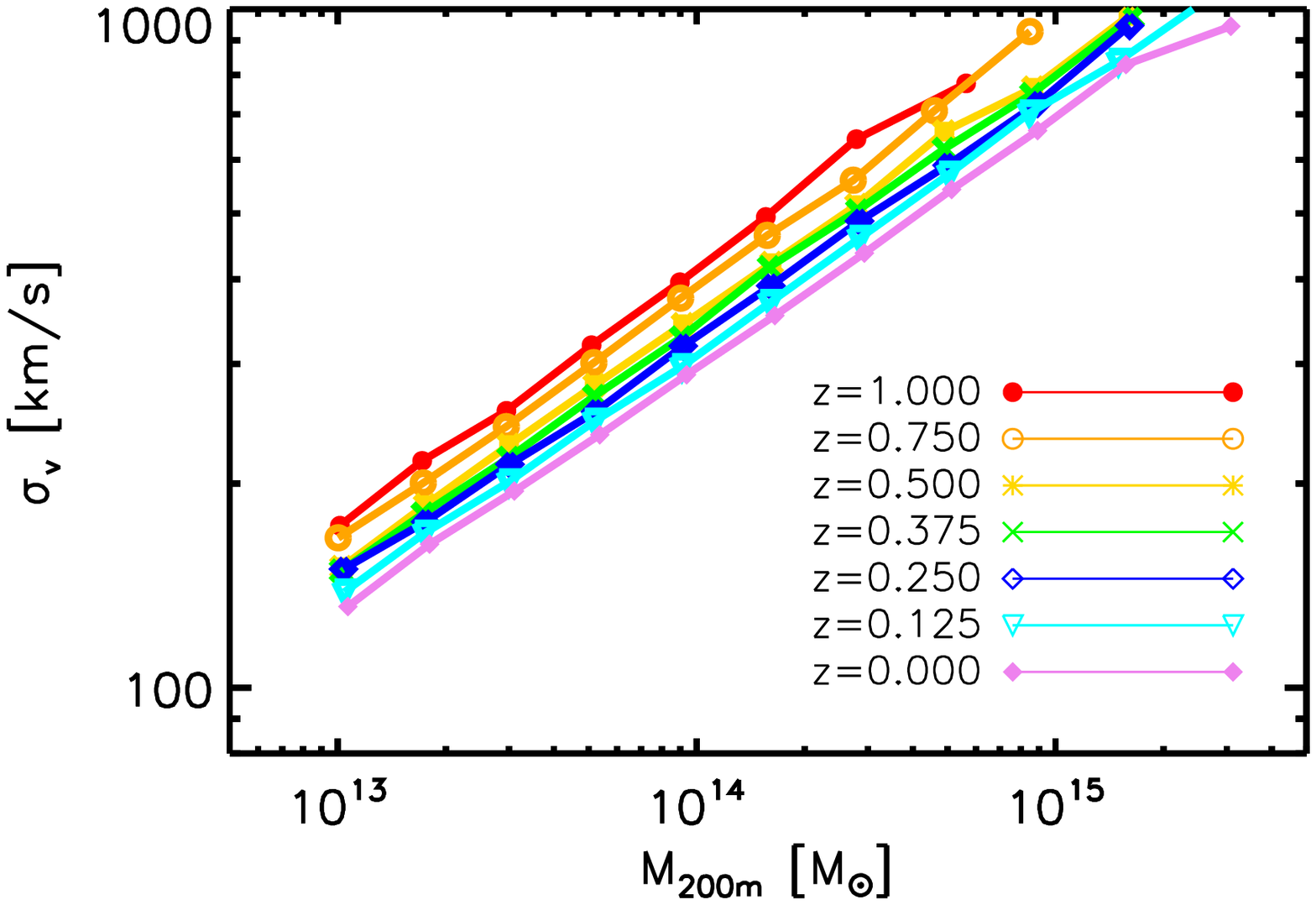}
\includegraphics[width=0.995\columnwidth]{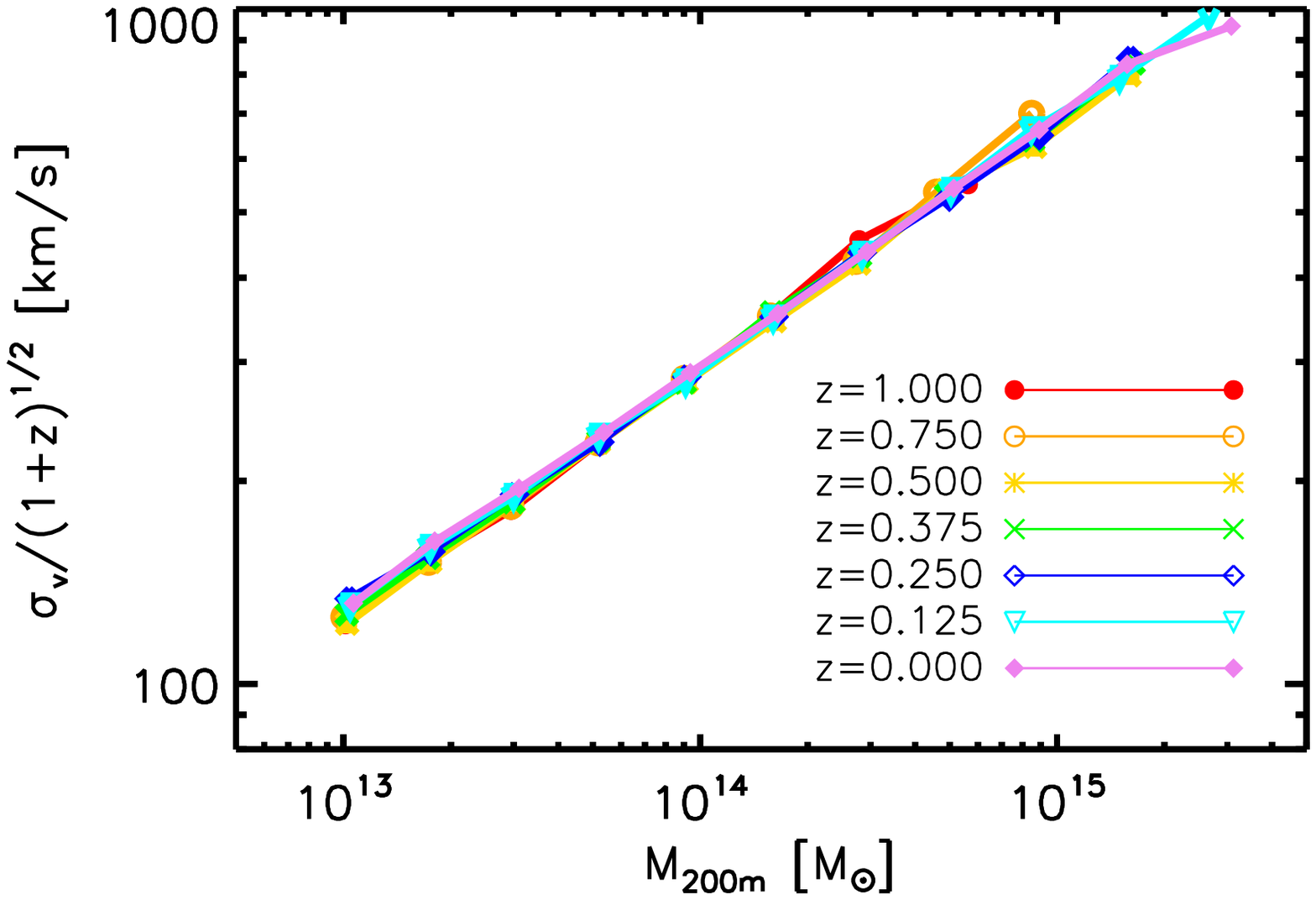}
\caption[]{Evolution of the mean $\sigma_v$--halo mass relation back to $z=1$. Velocity dispersions are calculated using member galaxies within a 3D radius, $r_{200m}$, and that have stellar masses $M_* \ge 10^{10} \rm M_\odot$ .  In the left panel we show the unscaled relations, while in the right panel the mean velocity dispersions have been re-scaled to account for self-similar evolution.  The velocity dispersion--halo mass relation evolves at the self-similar rate to a high level of accuracy.}
\label{fig:zevol}
\end{figure*}

To test the sensitivity of the velocity dispersions to baryon physics, we compare our (WMAP9) hydro simulation-based results with that derived from a dark matter only version of the simulation (i.e., using identical initial conditions but simulated with collisionless dynamics only).
To make a fair comparison with the dark matter only simulation, we should select (approximately) the same satellite population as in the hydro simulations.  In order to do this, we first assign stellar masses to the subhaloes using the subhalo abundance matching (SHAM) results of \citet{Moster13}.  Specifically, we convert the Moster et al. stellar mass--halo mass relation (including their estimated level of intrinsic scatter) into a stellar mass--maximum circular velocity ($V_{\rm max}$) relation, using the $M_{200}$--$V_{\rm max}$ relation for centrals from the dark matter simulation.  We then estimate the stellar masses of all subhaloes (centrals and satellites) using this stellar mass--$V_{\rm max}$ relation.  (We have explicitly checked that the resulting galaxy stellar mass function from our dark matter simulation reproduces the observed SDSS galaxy stellar mass function well, as found in \citealt{Moster13}.)  Furnished with stellar mass estimates for the subhaloes, we then apply the same galaxy and group selection criteria on the dark matter only simulation as imposed on the hydro simulations (as described in Section \ref{sec:selection}) and estimate the velocity dispersions in the same way.  We then match groups/clusters in the dark matter only simulation to those in the hydro simulation using the dark matter particle IDs.

In Fig. \ref{fig:simcomp} we compare the mean fractional difference in the velocity dispersions between the hydro and the dark matter only simulations, plotted as a function of the dark matter only halo mass.  For comparison, we also show the effect of baryon physics on the halo mass.  Baryon physics (AGN feedback, in particular) lowers the halo masses of galaxy groups by $\sim10\%$ (consistent with \citealt{Velliscig14}) and also reduces the velocity dispersions by $\approx5\%$, approximately independent of (the dark matter only) mass.  Comparing these differences to the differences in the predicted VDFs for different cosmological models (see Fig.~\ref{fig:vdf_data}), the effect is not large but is also not negligible.  Therefore, if one plans to use velocity dispersions from dark matter only simulations (+SHAM) to predict the VDF, the velocity dispersions should be appropriately scaled down by $\approx5\%$.  Alternatively, if one starts from a halo mass function from a dark matter only simulation, the halo masses first need to be adjusted (e.g., as proposed by \citealt{Velliscig14}) and then our hydro-simulated velocity dispersion--halo mass relation can be applied (including the scatter and evolution, as described below).

\subsubsection{Evolution}

To predict the evolution of the velocity dispersion counts we need to know how the velocity dispersion--halo mass relation evolves with redshift.  Under the assumption of self-similar evolution, the typical orbital velocity of a halo of fixed spherical over-density mass evolves as $\sigma_v \propto E(z)^{1/3}$, where $E(z)=[\Omega_m(1+z)^3 + \Omega_{\Lambda}]^{1/2}$, if the mass is defined with respect to the critical density, or as $\sigma_v \propto (1+z)^{1/2}$ if the mass is defined with respect to the mean matter density.  Note that even though we have already shown that the dependence on halo mass (the power-law index) at $z=0$ is not exactly self-similar, this does not automatically imply that the redshift evolution of the amplitude will not be well approximated with a self-similar scaling.  Indeed, such behaviour is seen in other variables such as the X-ray luminosity--temperature relation, which displays a strong departure from self-similarity in the slope of the relation but, according to some current analyses, evolves at a close to self-similar rate (e.g., \citealt{Maughan12}).

In the left panel of Fig.~\ref{fig:zevol} we plot the mean velocity dispersion--halo mass relation at a variety of redshifts going back to $z=1$.  Clearly there is a strong increase in the amplitude of the relation with increasing redshift.  In the right panel of Fig.~\ref{fig:zevol} we scale out the self-similar expectation, which has the effect of virtually removing the entire redshift dependence seen in the left panel.  In other words, to a high level of accuracy ($\lesssim 2 \%$) we find that the velocity dispersion--halo mass relation evolves self-similarly.   This statement remains the case if one instead defines the mass according to the critical density and uses $E(z)^{1/3}$ as the self-similar expectation, as opposed to $(1+z)^{1/2}$, so that:

\begin{equation}
\begin{split}
\sigma_v(M_{\rm \Delta,mean},z) & = \sigma_v(M_{\rm \Delta,mean},z=0) \ (1+z)^{1/2} \ \ \ {\rm or}, \\
\sigma_v(M_{\rm \Delta,crit},z) & = \sigma_v(M_{\rm \Delta,crit},z=0) \ E(z)^{1/3} .
\end{split}
\end{equation}

\subsubsection{Total scatter and its evolution}
\label{sec:scatter}
The scatter about the mean $\sigma_v$--halo mass relation is non-negligible at all masses and can be particularly large at low masses, due to poor sampling (as we will show below).  Modelling this scatter is necessary if one wishes to predict the velocity dispersion counts by convolving the velocity dispersion--halo mass relation with a halo mass function, as Eddington bias will become quite important.  Here we characterise the scatter in the velocity dispersion as a function of halo mass and redshift.

\begin{figure*}
\includegraphics[width=1.7\columnwidth]{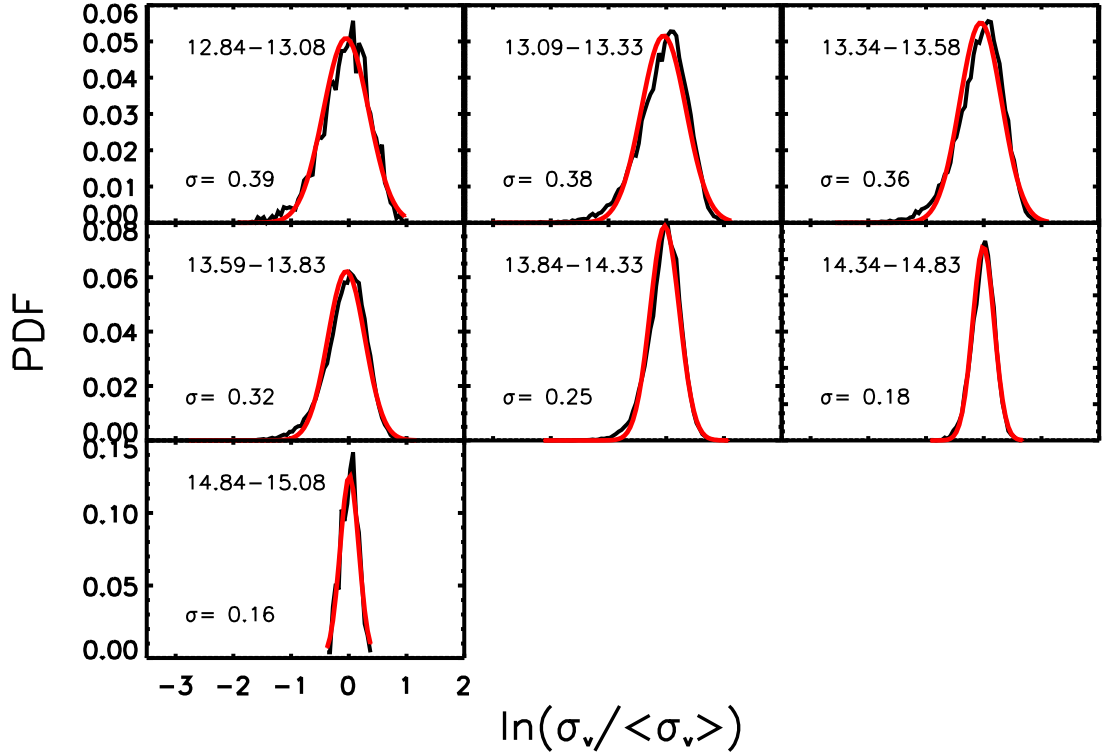}
\caption[]{Residuals about the best-fit power-law to the mean velocity dispersion--halo mass relation.  The 7 histograms correspond to different mass bins.  The solid black curve represents the residuals about the mean, while the solid red curve represents the best-fit lognormal distribution.    To boost our statistics, we stack the velocity dispersions from all redshifts and vary the binning in halo mass.  Lognormal distributions describe the residuals about the mean relation quite well, but the width of the distribution (i.e., the scatter) about the mean decreases strongly with halo mass.}
\label{fig:histofit}
\end{figure*}

To aid our analysis of the scatter, we first divide the velocity dispersion of each system by that predicted by the best-fit power-law to our mean velocity dispersion--halo mass relation.  After dividing out the mean mass relation, the residuals (see Fig.~\ref{fig:histofit}) clearly show that the scatter decreases with mass. To improve statistics, the velocity dispersions for different redshifts have been rescaled to $z=0$ using Eqn.\ 6, stacked, and binned to model the scatter as function of halo mass. The bin widths are chosen to equally sample the range in log$_{10}$ halo mass space, while avoiding large statistical errors from low bin populations. The first four halo mass bins are 0.25 dex in width, increasing to 0.5 dex for the following two bins, and final bin has a width of 0.25 dex. 

It is interesting to note that previous studies that used dark matter particles or subhaloes to estimate the velocity dispersions (e.g., \citealt{Evrard08,Munari13}) found that the scatter did not vary significantly with system mass.  The difference between these works and the current one is that we select only relatively massive galaxies, which should be more appropriate for comparisons to observations.  Since massive galaxies become increasingly rare in low-mass groups, the {\it statistical} uncertainty in the derived velocity dispersion increases.  Studies that use dark matter particles (or, to a lesser extent, all dark matter subhaloes), on the other hand, have essentially no statistical error and therefore any scatter present is likely to be intrinsic in nature (e.g., due to differences in state of relaxation).  These studies therefore suggest that the intrinsic scatter does not depend significantly on halo mass, a finding which we confirm below.

We fit the total scatter residuals about the mean relation in each mass bin with a lognormal distribution.  Fig.~\ref{fig:histofit} shows histograms of logged velocity dispersion residuals, and the normal curve fit.  A lognormal distribution describes the residuals well in all of the mass bins we consider.  We note that in the first three (lowest) mass bins, the distribution becomes somewhat skewed relative to lognormal when systems with less than 5 members are included in the analysis.  As discussed in Section 3.1, we have excluded these systems from our analysis, noting that when comparing to observed velocity dispersion counts from GAMA (Caldwell et al., in prep), we also plan to impose a richness cut of $\ge 5$ on the observed sample.  

\begin{figure}
\includegraphics[width=0.995\columnwidth]{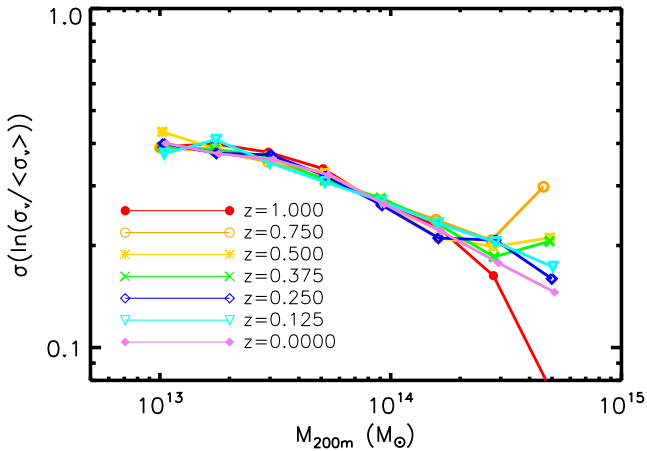}
\caption[]{Evolution of the total scatter about the mean velocity dispersion--halo mass relation for seven redshifts from $z=0$ to $z=1$.  There is no evidence for significant evolution in the scatter about the mean relation. }
\label{fig:zslns}
\end{figure}

In Fig.~\ref{fig:zslns} we show the evolution of the total scatter--halo mass relation for seven redshifts from $z=0$ to $z=1$.  Here one can more clearly see that the scatter varies strongly with halo mass.  However, it does not appear to vary significantly with redshift, at least back to $z=1$. 

\subsubsection{Decomposing the total scatter into statistical and intrinsic components}

Although quantifying the total scatter as a function of halo mass (in order to interpolate it with a halo mass function later) is the primary goal of this section, a deeper understanding of the scatter is required if we wish to consistently compare with observations.  That is because the scatter is composed of both intrinsic and statistical components and the latter is clearly going to be a function of observational survey parameters (e.g., limiting magnitude).  We therefore proceed to decompose the total scatter into its two components. 

We have focused so far on the (total) scatter as a function of mass, but the statistical component is best understood through its dependence on richness, since fundamentally it is the number of tracers that determines how well the (true) velocity dispersion can be determined.

Statistical scatter is the scatter caused by randomly sampling a distribution with a finite number of points.  In our particular case, sampling the velocity distribution of a galaxy group or cluster with a finite number of galaxies means that we can only measure the velocity dispersion to a certain level of accuracy.  Clearly, the more tracer galaxies we have, the more precise and accurate our measurement of the velocity dispersion will become.

To help understand the level of statistical scatter contributing to the total scatter, we use simple Monte Carlo simulations to determine the accuracy to which the velocity dispersion of a system can be determined given a finite number of tracers.  We assume a normal distribution for the velocities and vary the number of tracers from 2 up to 1500 (which approximately spans the range of richnesses relevant for groups and clusters), drawing 1000 random samples for each number of tracers we consider.  So, for example, to determine how well one can measure the velocity dispersion for a system with 5 members, we would randomly draw 5 velocities from a normal distribution and then compute the velocity dispersion using the gapper method.  We repeat this 1000 times, each time recording the derived velocity dispersion.  This gives us a spread of velocity dispersions at fixed richness, which we then fit with a lognormal distribution.  The width of this lognormal distribution is the statistical scatter in the velocity dispersion for a system with 5 members. 

\begin{figure}
\includegraphics[width=0.995\columnwidth]{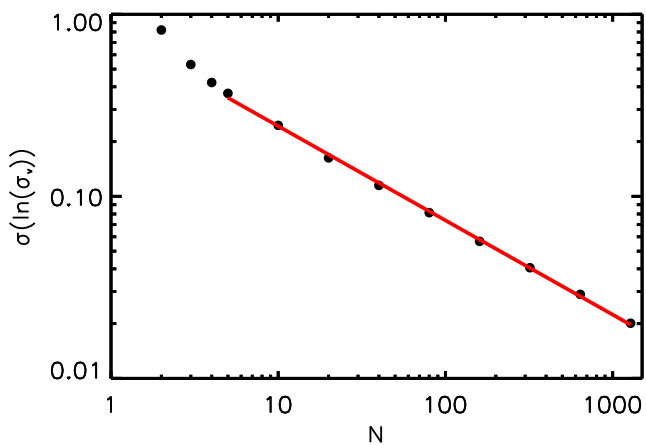}
\caption[]{Statistical scatter as a function sample size, $N$, determined from Monte Carlo (MC) simulations (see text).  The black points are the calculated value (derived from the MC simulations) for each sample size, and the red line is a power-law fit to the points with $N\ge5$. A simple power-law relation works well for $N\ge5$.  }
\label{fig:scattervngal}
\end{figure}

In Fig.~\ref{fig:scattervngal} we plot the derived statistical scatter as a function of richness.  As expected, the statistical scatter increases with decreasing richness.  We find that for $N \ge 5$, the scatter is well modelled by a simple power-law of the form:

 \begin{equation}
\sigma_{\rm stat}(\ln(\sigma_v)) = 0.07\Big(\frac{N}{100}\Big)^{-0.5} {\rm for} \ N \ge 5
\label{eq:plainstat}
\end{equation}

This result is generally applicable for systems that have an underlying normal distribution, regardless of whether they are simulated or real clusters. Note that this does not depend on whether the multiplicative Eke correction is applied because the scatter is modelled in $\ln(\sigma_v)$.

We now have a measurement of the statistical scatter at fixed richness.  In analogy to Fig.~\ref{fig:zslns}, we can compute the {\it total} scatter in bins of richness as opposed to mass (i.e., we compute the scatter in the residuals about the mean velocity dispersion--richness relation).  The total scatter is just composed of statistical and intrinsic components (summed in quadrature), so we can now also determine the intrinsic scatter as a function of richness.

\begin{figure}
\includegraphics[width=0.995\columnwidth]{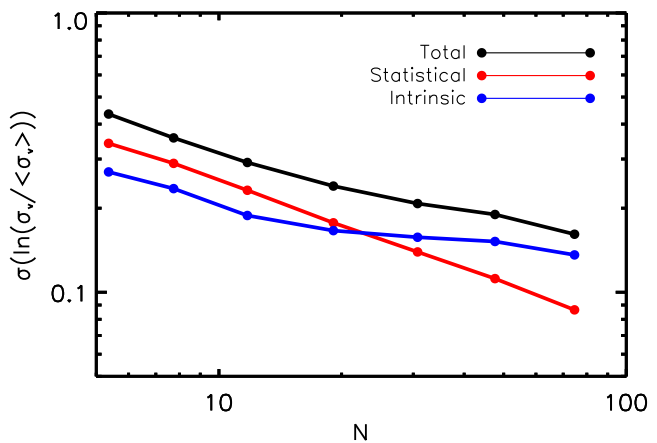}
\caption[]{Contributions of intrinsic and statistical scatter to the total scatter about the mean velocity dispersion--richness relation, for the case of a \planck~cosmology with massless neutrinos and selecting only groups with at least 5 member galaxies with stellar masses $M_* \ge 10^{10} \rm M_\odot$ and that are within $r_{200m}$.  The black curve is the total scatter, the red curve is the statistical scatter, and the dashed blue curve is the derived intrinsic scatter (assuming the intrinsic and statistical scatters sum in quadrature to give the total scatter).  Statistical scatter dominates for all but the most rich/massive systems.  The intrinsic scatter does not depend strongly on richness/mass. }
\label{fig:scatter_components}
\end{figure}

In Fig.~\ref{fig:scatter_components} we show the contribution of the statistical and intrinsic scatter to the total scatter as a function of richness.  We find that statistical scatter dominates the total scatter for all but the richest (highest-mass) systems.  

Note that it is galaxy selection criteria that determines the degree of statisical scatter.  In the simulations we use a galaxy stellar mass limit of $10^{10}$M$_\odot$, but if we were able to lower that limit (e.g., by using higher resolution simulations) the statistical scatter would decrease.  Likewise for observational surveys, if the apparent magnitude limit of the survey were increased (i.e., so that we could measure fainter systems), the number of galaxies will increase and so too will the accuracy of the velocity dispersions.  Other selection criteria (such as red sequence selection) can also affect the estimated velocity dispersion (e.g., \citealt{Saro13}) via their influence on the number of tracers used to measure the velocity dispersion.

Note that while the statistical scatter is a strong function of richness, the intrinsic scatter does not vary significantly over the range of richnesses we have examined, consistent with previous studies (e.g., \citealt{Evrard08,Munari13}). In Appendix A, we provide the mean intrinsic scatter for a variety of mass definitions and apertures. The average intrinsic scatter varies little with mass definition and choice of aperture with values $\approx$ 0.19 dex in $ln \sigma$.

\subsection{Summary of velocity dispersion--mass relation}
\label{sec:method_summary}
Here we summarize our characterisation of the velocity dispersion--halo mass relation for groups with at least 5 members with stellar masses $M_* \ge 10^{10} \rm M_\odot$.  The mean relation can be well-described by a simple power-law spanning low-mass groups to high-mass clusters (see Fig.~\ref{fig:msmatch}) is approximately independent of cosmology (for example, the amplitude for the mean $\sigma_v-\rm M$ power-law differs by $\approx0.3\%$ between Planck and WMAP9 cosmologies). The mean power-law evolves self-similarly back to $z=1$ at least (see Fig.~\ref{fig:zevol}).  Note that the amplitude of the relation is $\approx5\%$ lower than that predicted by a dark matter only simulation where a consistent selection of satellites is applied (see Fig.~\ref{fig:simcomp}).  The scatter about the mean relation can be well-represented by a lognormal distribution whose width varies strongly as a function of halo mass (see Fig.~\ref{fig:histofit}) but not with redshift (see Fig.~\ref{fig:zslns}).  The strong variation in scatter with halo mass is due to the increasing importance of statistical scatter with decreasing mass/richness (see Fig.~\ref{fig:scattervngal}), whereas the intrinsic scatter does not depend significantly upon mass/richness and is only important for systems with richnesses exceeding several tens (see Fig.~\ref{fig:scatter_components}).


\subsection{Testing the model}
\label{sec:test_model}
We now test the accuracy of our simple velocity dispersion--halo mass relation model by convolving it with the halo mass distribution drawn from the simulations and comparing the predicted velocity dispersion distribution with the one drawn directly from the simulations.  In particular, for the model prediction, we use the mass of each halo to infer the predicted mean velocity dispersion using eqns.~(3) and (6).  We then (additively) apply scatter by randomly drawing from a lognormal distribution with a width set by the total scatter--halo mass relation, which we characterise with the black curve in Figure \ref{fig:scatter_components}.

\begin{figure}
\includegraphics[width=0.995\columnwidth]{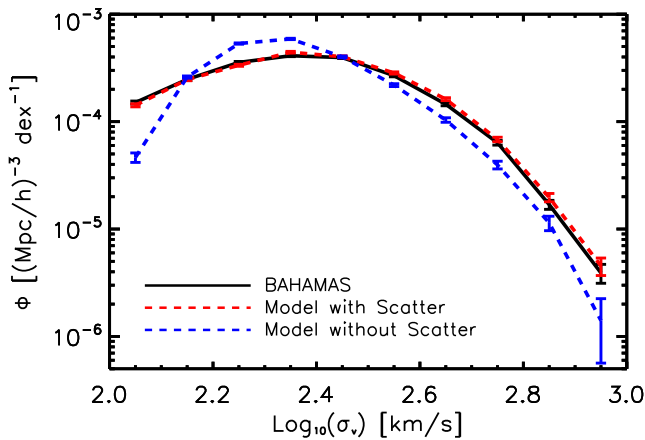}
\caption[]{Comparison of the VDF from the \planck~2013 (massless neutrino) simulation (solid black curve) with that predicted by a simple model of the velocity dispersion--halo mass relation convolved with the halo mass distribution from the simulations (red dashed curve).  Also shown is the model prediction when the scatter in the velocity dispersion--halo mass is ignored (blue dashed curve).  The model with scatter reproduces the simulation VDF quite well over the full range of velocity dispersions.  Ignoring the effects of scatter and associated Eddington bias leads to an underestimate of the number of systems with velocity dispersions exceeding 300 \kms.}
\label{fig:vdf_match}
\end{figure}

Figure \ref{fig:vdf_match} compares the VDF derived directly from the simulations with that predicted by our simple model of the velocity dispersion--halo mass relation convolved with the halo mass distribution, both imposing a richness cut of $N \ge 5$.  We also show the effect of ignoring the scatter in the velocity dispersion--halo mass relation.  In spite of its simplicity, the model prediction (with scatter) reproduces the simulation VDF remarkably well (to better than 10-15\% accuracy) over the full range of velocity dispersions that we sample.  By contrast, ignoring the scatter causes the curve to strongly under predict the VDF above velocity dispersions of 300 \kms.  Modelling the scatter is therefore crucially important if one wishes to make an accurate prediction for the velocity dispersion counts and obtain unbiased constraints on cosmological parameters.

\begin{figure}
\includegraphics[width=0.995\columnwidth]{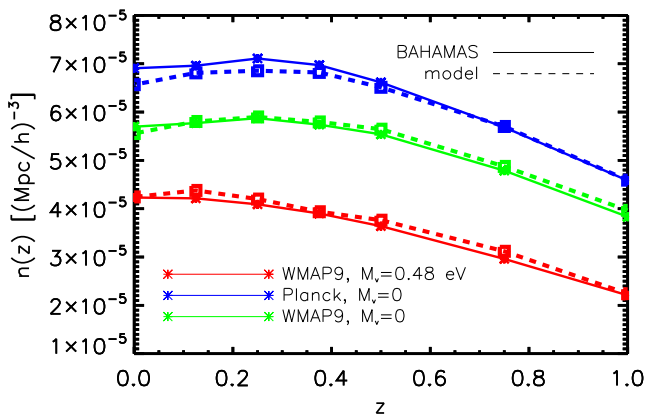}
\caption[]{The number density of systems with $\sigma_v\ge300$ km/s as a function of redshift. Solid lines are from the simulation, dashed lines are velocity dispersions constructed from the models described in the previous section and convolved with the halo mass function from the BAHAMAS simulation. The colours indicate different cosmologies: blue=Planck, green= \wmap, and red = WMAP9 with neutrino mass = 0.48 eV. }
\label{fig:nz_match}
\end{figure}

In Figure \ref{fig:nz_match}, we compare the evolution of the velocity dispersions counts for systems with $\sigma_v \ge 300$ km/s from various simulations with different cosmologies with that predicted by our simple model.  There is good agreement with between model predictions and the simulations.  

Finally, we note that in the above analysis the effects of feedback have already implicitly been included.  As demonstrated in Section 4.1.2, feedback can affect both the halo mass and the velocity dispersion.  Therefore, in order to predict the velocity dispersion function from the halo mass function one must appropriately account for feedback effects on the halo mass and then apply the above velocity dispersion--halo mass relation.  The modification of the halo masses are already implicitly included in our analysis, as we use the halo mass distribution directly from the hydro simulations.  If, however, one wishes to use theoretical mass functions in the literature that are based dark matter simulations (e.g., \citealt{Sheth01,Tinker08}) appropriate feedback modifications should be applied (such as those proposed by Velliscig et al.). 

\section{Cosmological constraint forecasts}
\label{sec:constraints}

In Section 4 we outlined a simple yet accurate method for predicting the velocity dispersion counts for different cosmologies.  Here we use this apparatus to make some simple forecasts for current and future spectroscopic surveys.  In particular, we examine the kind of constraints that these surveys will place on the $\sigma_8$--$\Omega_m$ plane and on the summed mass of neutrinos.  

We consider three different synthetic spectroscopic surveys, with characteristics chosen to approximately match those of the completed GAMA survey \citep{Driver11}, the upcoming WAVES-Wide survey \citep{Driver15}, and the upcoming DESI bright galaxy survey \citep{Levi13}.  For the synthetic GAMA-like survey, we adopt a survey field of view of 180 square degrees and galaxy stellar mass limit of $10^{10} \rm{M}_{ \odot}$.  For the synthetic WAVES-like survey, we adopt 1000 square degress and a stellar mass limit of $10^{9} \rm{M}_{ \odot}$.  For the synthetic DESI-like survey, we adopt 14,000 square degrees and a stellar mass limit of $10^{10} \rm{M}_{ \odot}$.  For all three cases we examine the cosmological constraints that can be derived using the velocity dispersion number counts exceeding 300 \kms\ within a redshift $z<0.2$.  We note that it may be possible to obtain improved constraints by looking at multiple thresholds in velocity dispersion and/or multiple redshift bins, which we intend to explore further in future work. 




\subsection{$\sigma_8$--$\Omega_m$ plane}

We construct a 151$\times$151 grid of [$\sigma_8$,$\Omega_m$] values ranging from $0.7<\sigma_8<0.9$ and $0.2<\Omega_m<0.4$.  For the other parameters, we adopt a `WMAP9-based' cosmology, fixing $h=0.7$, $\Omega_b=0.0463$, $n_s=0.972$ and $\Omega_\Lambda=1-\Omega_m$.  For a given set of cosmological parameters (of which there are 22801 independent sets), we use {\textsc{camb}} to compute the $z=0$ linear transfer function, which is used as input for the \citet{Tinker08} halo mass function.  We convolve the predicted halo mass function with the halo mass--velocity dispersion relation derived in the previous sections.  Note that for the case of the synthetic WAVES-like survey, we have decreased the statistical scatter in the velocity dispersions in line with the adopted lower stellar mass limit of that survey.  This was done by using the abundance matching procedure described in Section 4.1.2 to estimate how much the richnesses would increase by dropping the stellar mass limit from $10^{10} \rm{M}_{ \odot}$ to $10^{9} \rm{M}_{ \odot}$.

\begin{figure}
\includegraphics[width=0.995\columnwidth]{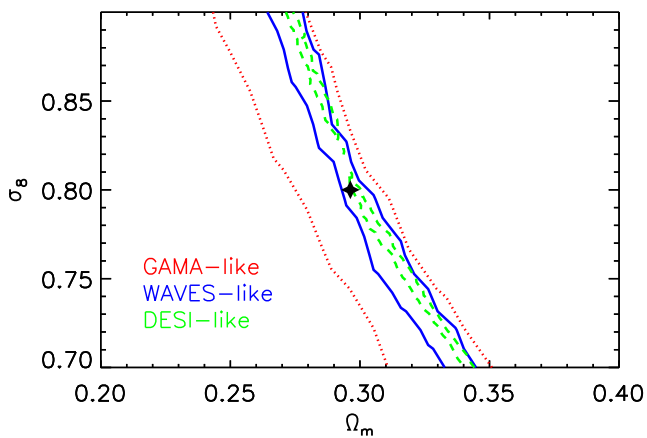}
\caption[]{Forecasted constraints on $\sigma_8$ and $\Omega_m$ using the velocity dispersion number counts.  Dashed contours define the 1-$\sigma$ confidence interval for the GAMA-like, WAVES-like, and DESI-like synthetic surveys that we consider.  The black star indicates the adopted test cosmology.  The joint constraint scales approximately as $\sigma_8 \Omega_m$ (see text).  The amplitude can be determined to approximately $20\%$, $10\%$ and $4\%$ accuracy with the GAMA-like, WAVES-like, and DESI-like synthetic surveys, respectively.}
\label{fig:s8om}
\end{figure}

Figure \ref{fig:s8om} shows the 1-$\sigma$ confidence interval for a test cosmology of $\sigma_8=0.8$ and $\Omega_m=0.3$; i.e., we assume these are the truth and see how well this is recovered.  The 1-$\sigma$ confidence interval shows a strong degeneracy in the joint constraints on $\sigma_8$ and $\Omega_m$, as expected.  We find that a simple power law with $\sigma_8 \propto \Omega_m^\alpha$ with $\alpha \approx -1$ describes the degeneracy relatively well.  The exact slope of the degeneracy depends somewhat on which synthetic survey is considered; we find $\alpha=-0.86\pm0.01$, $-1.08\pm0.01$, and $-1.13\pm0.01$ for the GAMA-like, WAVES-like, and DESI-like surveys, respectively.

It is worth noting that the degeneracy found here is significantly steeper than that found in some previous halo mass counts studies, which indicate $\alpha \approx -0.6$ (e.g.,\citealt{Vikhlinin09,Rozo10}).  The reason for this difference is not that we are using velocity dispersions as opposed to halo mass, but is instead due to the specific velocity dispersion threshold of 300 \kms\ that we adopt.  In particular, this velocity dispersion threshold corresponds roughly to a halo mass of $\sim 10^{14} {\rm M}_\odot$, which is lower than most current halo mass counts studies (certainly compared to X-ray- and SZ-based studies).  Note that the abundance of groups is somewhat more sensitive to $\Omega_m$ than to $\sigma_8$, whereas the reverse is true for high-mass clusters.  We have verified that using higher velocity dispersion thresholds leads to a flatter degeneracy between $\sigma_8$ and $\Omega_m$, similar in shape to that found previously for studies based on massive clusters.  This motivates our comment above, that one can potentially use multiple velocity disperion thresholds to help break the degeneracy between the two cosmological parameters.

It is immediately evident from Figure \ref{fig:s8om} that upcoming spectrscopic surveys will severely constrain the amplitude of the degeneracy.  We can quantify this by comparing the width of the 1-$\sigma$ confidence interval (i.e., the width perpendicular to the degeneracy) to the best-fit amplitude.  We find that a GAMA-like survey would be expected to constrain the amplitude to $\approx20\%$, whereas a WAVES-like survey would constrain it to $\approx10\%$ and a DESI-like survey would constrain it to better than $4\%$ accuracy.

Note that in the above analysis we have held the other cosmological parameters fixed.  Allowing these to be free will likely broaden the constraints on $\sigma_8$ and $\Omega_m$ slightly.

\begin{figure}
\includegraphics[width=0.995\columnwidth]{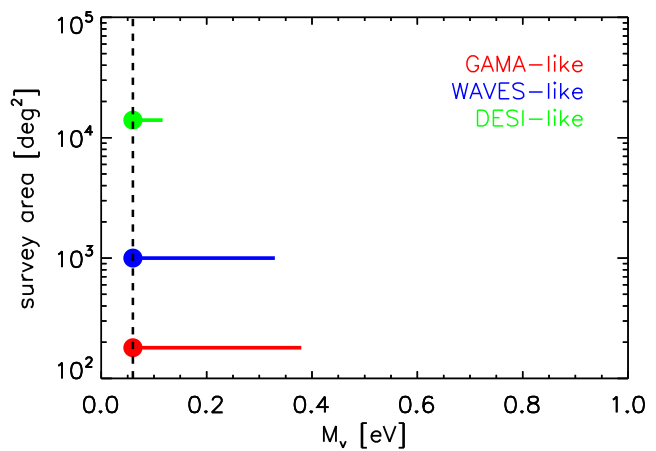}
\caption[]{Forecasted constraints on the summed mass of neutrinos, $M_\nu$.  The 1-$\sigma$ confidence intervals are plotted in red, blue and green for the GAMA-like, WAVES-like, and DESI-like synthetic surveys that we consider.  We adopted $M_\nu=0.06$ eV as the test cosmology.  }
\label{fig:numass}
\end{figure}

\subsection{Summed mass of neutrinos, $M_\nu$}

Here we examine how well the velocity dispersion counts can be used to constrain the summed mass of neutrinos.  For this case we adopt a \planck-based cosmology, fixing $h=0.6726$, $\Omega_b=0.0491$, $\Omega_{\rm cdm}=0.2650$, $n_s=0.9652$, and assume a flat universe (i.e., as we increase $M_\nu$ and $\Omega_\nu$, $\Omega_\Lambda$ is decreased to maintain $\Omega_{\rm tot}=1$).  By holding all parameters apart from $M_\nu$ and $\Omega_\Lambda$ fixed, we are essentially considering a case where we take the primary CMB cosmology to be a correct description of the Universe at early times and quantify how well adding measurements of the velocity dispersion counts constrains the summed mass of neutrinos.

We consider 151 different values of the summed neutrino mass, ranging from the minimum allowed value of 0.06 eV up to 1 eV.  We adopt $M_\nu=0.06$ eV as our test case.

In Fig. \ref{fig:numass} we explore the constraining power of the three synthetic surveys described above.  The error bars show the 1-$\sigma$ confidence errors.  A GAMA-like survey, when combined with primary CMB constraints, would be expected to constrain $M_\nu \la 0.38$ eV.  A WAVES-like survey will improve on this somewhat, while a DESI-like survey will tightly constrain the summed mass of neutrinos ($M_\nu < 0.12$ eV) when it is combined with primary CMB measurements.  The potential constraints from a DESI-like experiment are interesting from a particle physics perspective, as they could potentially allow one to distinguish between the `normal' and `inverted' neutrino hierarchy scenarios (see \citealt{Lesgourgues06} for further discussion).  However, we note that our forecasts are still fairly simplistic, in that we have held the other cosmological parameters fixed (although they are strongly constrained by the primary CMB) and we have not considered the effects of redshift errors, group selection, etc.  On the other hand, we have also not used the full information available in our dataset (e.g., multiple redshifts and velocity dispersion thresholds), which would be expected to improve the precision of the constraints.

\section{Discussion and Conclusions}

Recent work has highlighted the importance of systematic uncertainties in halo mass measurements for `cluster cosmology'.  Motivated by this, we have proposed an alternative test, which is the number counts as a function of one-dimensional velocity dispersion of the galaxies in the cluster (the VDF), as opposed to halo mass.  We argue that the velocity dispersion can be predicted basically as robustly as the mass in cosmological simulations but, unlike the mass, the velocity dispersion can be directly observed, thus offering a way to make a direct comparison of cluster counts between theory and observations.  We note that the proposed use of velocity dispersion counts to probe cosmology is not new.  In pioneering work, \citet{Evrard08} previously used dark matter only simulations to show that one could constrain the amplitude of density fluctuations ($\sigma_8$) in this way.  Here we have extended these ideas and applied them to realistic hydrodynamical simulations.

We have used the \calsim~suite of cosmological hydrodynamical simulations \citep{McCarthy16} to explore the cosmological dependence of the VDF, which we also find to be strong (see Figs.~\ref{fig:vdf_data} and \ref{fig:nz_data}).  For example, at a velocity dispersion $\sigma_v\approx1000$ km/s, adopting a \planck~2013 cosmology results in $\approx50\%$ more systems compared to adopting a \wmap~cosmology (both assuming massless neutrinos).  Even at a relatively modest velocity dispersion of $\approx 300$ km/s (corresponding to haloes with masses $\sim10^{14} \rm M_\odot$) the difference is still significant ($\approx 20\%$).  The addition of a massive neutrino component strongly suppresses the number of high-velocity dispersion systems, as expected.

Unfortunately, the expense of large-scale simulations like \calsim~prohibits us from fully sampling the full range of cosmological parameters allowed by current experiments.  Therefore, to place robust constraints on cosmological parameters using the VDF requires a method to quickly compute the predicted VDF for a given set of parameters.  We have proposed a simple method to achieve this goal: convolution of the simulation-based velocity dispersion--halo mass relation with theoretically predicted halo mass functions, that have been appropriately modified to take into account feedback (e.g., \citealt{Velliscig14}).  

We have shown that the mean relation is well-characterised by a simple power-law spanning low-mass ($\approx 10^{12.7} \rm M_{\odot}$) groups to high-mass ($\approx 10^{15} \rm M_{\odot}$) clusters (see Fig.~\ref{fig:msmatch}) which evolves according to the self-similar expectation (see Fig.~\ref{fig:zevol}) and does not depend significantly on cosmology (see Fig.~\ref{fig:simcomp}).  Note that the amplitude of the relation is $\approx5\%$ lower than that predicted by a dark matter only simulation where a consistent selection of satellites is applied (see Fig.~\ref{fig:simcomp}).  The scatter about the mean relation is lognormal with a width that varies strongly as a function of halo mass (see Fig.~\ref{fig:histofit}) but does not vary with redshift (see Fig.~\ref{fig:zslns}).  The strong variation in scatter with halo mass is due to the increasing importance of statistical scatter at low masses due purely to decreasing richness (see Fig.~\ref{fig:scattervngal}), whereas the intrinsic scatter does not depend significantly upon mass/richness and only becomes important for systems with several tens of galaxies (see Fig.~\ref{fig:scatter_components}).  We have shown that, in spite of the simplicity of our model for the velocity dispersion--halo mass relation, it recovers the VDF and number counts derived directly from the simulation quite well (see Figs.~\ref{fig:vdf_match} and \ref{fig:nz_match}).  

In Section 5 we demonstrated that measurements of the velocity dispersion counts with current spectroscopic surveys such as GAMA, and (especially) with upcoming wide-field surveys such as WAVES and DESI, can be used to strongly constrain the $\sigma_8$--$\Omega_m$ plane (Fig.~\ref{fig:s8om}) and, when combined with primary CMB measurements, the summed mass of neutrinos (Fig.~\ref{fig:numass}).

Finally, in the present study we have made predictions for an essentially perfect observational survey, where all groups above a given velocity dispersion and richness cut are accounted for and with zero contamination (i.e., false positives).  Clearly, these conditions are never strictly met in real observational surveys.  To address these issues we advocate the use of synthetic (mock) surveys, which can be analysed in the same way as the data.  This allows one to implicitly include the effects of completeness and impurity in the predictions, and it also ensures similar statistical scatter.  In a follow up paper (Caldwell et al., in prep), we plan to compare our theoretical predictions to the GAMA galaxy group catalog \citep{Robotham11} using such synthetic surveys.

\section*{Acknowledgments}
We thank the anonymous referee for helpful suggestions that improved the quality of the paper. IGM acknowledges support from a STFC Advanced Fellowship. SB was supported by NASA through Einstein Postdoctoral Fellowship Award Number PF5-160133. JS acknowledges support from ERC grant 278594 -- GasAroundGalaxies.

This work used the DiRAC Data Centric system at Durham University, operated by the Institute for Computational Cosmology on behalf of the STFC DiRAC HPC Facility (www.dirac.ac.uk). This equipment was funded by BIS National E-infrastructure capital grant ST/K00042X/1, STFC capital grants ST/H008519/1 and ST/K00087X/1, STFC DiRAC Operations grant ST/K003267/1 and Durham University. DiRAC is part of the National E-Infrastructure.

\bibliographystyle{mnras}
\bibliography{mybib}

\appendix

\section{Velocity dispersion--halo mass relations for alternative mass definitions and apertures}
In Table \ref{table:coeffs} we present models for the velocity dispersion- mass relation and its scatter. Since the relation changes slightly depending on the distribution of the galaxies in the cluster, we have calculated the fits for several mean and critical mass definitions and cluster radii.

\begin{table*}

\begin{center}
\caption{Power-law fits to the $z=0$ $\sigma_v$--halo mass relation for \planck~2013 cosmology.  Fits are of the form $\log_e(y)=a+b\log_e[M/10^{14} \rm M_\odot]$. The average intrinsic scatter is provided for each halo mass and aperture cut. The value for intrinsic scatter quoted below adds with the natural logarithm of statistical scatter in quadrature to equal the $\rm log_e$ of total scatter for a group or cluster on the velocity dispersion-halo mass plane. }
\begin{tabular}{|lllll}
Halo mass & Aperture &  $\sigma_v$--$M$ intercept & $\sigma_v$--$M$ slope & intrinsic scatter\\
\hline
\hline
$M_{500,mean}$    & $R_{500,mean}$     & 5.7788       & 0.4003    & 0.1881   \\
$M_{500,crit}$    & $R_{500,crit}$     & 6.0084       & 0.4113       & 0.1897\\
$M_{200,mean}$    & $R_{200,mean}$     & 5.6366       & 0.3852       &   0.1864  \\
$M_{200,crit}$    & $R_{200,crit}$     & 5.8220       & 0.4019       &  0.1906  \\
$M_{200,mean}$    & 1 Mpc            & 5.6672       & 0.3986       &  0.1877   \\
$M_{200,crit}$    & 1 Mpc            & 5.8138       & 0.3908       &  0.1877  \\
$M_{200,mean}$    & 0.5 Mpc          & 5.7104       & 0.4060       & 0.1889   \\
$M_{200,crit}$    & 0.5 Mpc          & 5.8583       & 0.4058       &  0.1889\\
\hline
\label{table:coeffs}
\end{tabular}
\end{center}
\end{table*}

\bsp	
\label{lastpage}
\end{document}